\documentclass{aastex631}

\usepackage{textcomp}
\usepackage[normalem]{ulem}

\usepackage{amsmath}

\usepackage[utf8]{inputenc}
\usepackage{tabularx}

\newcommand\clearrow{\global\let\rowmac\relax}

\usepackage{booktabs}
\usepackage{hyperref}
\usepackage{multirow}
\usepackage{amsmath}
\usepackage{graphicx} 

\shorttitle{Inferring Line-of-Sight Velocities and Doppler Widths from Stokes Profiles of 
GST/NIRIS}
\shortauthors{Jiang et al.}

\graphicspath{{./}{figures/}}

\begin{document}

\title{Inferring Line-of-Sight Velocities and Doppler Widths from Stokes Profiles of GST/NIRIS Using Stacked Deep Neural Networks}

\author{Haodi Jiang}
\affiliation{Institute for Space Weather Sciences, New Jersey Institute of Technology, University Heights, Newark, NJ 07102-1982, USA;
hj78@njit.edu, wangj@njit.edu, haimin.wang@njit.edu}
\affiliation{Department of Computer Science, New Jersey Institute of Technology, University Heights, Newark, NJ 07102-1982, USA}

\author{Qin Li}
\affiliation{Institute for Space Weather Sciences, New Jersey Institute of Technology, University Heights, Newark, NJ 07102-1982, USA;
hj78@njit.edu, wangj@njit.edu, haimin.wang@njit.edu}
\affiliation{Center for Solar-Terrestrial Research, New Jersey Institute of Technology, University Heights, Newark, NJ 07102-1982, USA}

\author{Yan Xu}
\affiliation{Institute for Space Weather Sciences, New Jersey Institute of Technology, University Heights, Newark, NJ 07102-1982, USA;
hj78@njit.edu, wangj@njit.edu, haimin.wang@njit.edu}
\affiliation{Center for Solar-Terrestrial Research, New Jersey Institute of Technology, University Heights, Newark, NJ 07102-1982, USA}
\affiliation{Big Bear Solar Observatory, New Jersey Institute of Technology, 40386 North Shore Lane, Big Bear City, CA 92314-9672, USA}

\author{Wynne Hsu}
\affiliation{Institute of Data Science,	National University of Singapore, 
Singapore 119077}
\affiliation{Department of Computer Science, School of Computing, National University of Singapore, Singapore 119077}

\author{Kwangsu Ahn}
\affiliation{Big Bear Solar Observatory, New Jersey Institute of Technology, 40386 North Shore Lane, Big Bear City, CA 92314-9672, USA}

\author{Wenda Cao}
\affiliation{Institute for Space Weather Sciences, New Jersey Institute of Technology, University Heights, Newark, NJ 07102-1982, USA;
	hj78@njit.edu, wangj@njit.edu, haimin.wang@njit.edu}
\affiliation{Center for Solar-Terrestrial Research, New Jersey Institute of Technology, University Heights, Newark, NJ 07102-1982, USA}
\affiliation{Big Bear Solar Observatory, New Jersey Institute of Technology, 40386 North Shore Lane, Big Bear City, CA 92314-9672, USA}

\author{Jason T. L. Wang}
\affiliation{Institute for Space Weather Sciences, New Jersey Institute of Technology, University Heights, Newark, NJ 07102-1982, USA;
hj78@njit.edu, wangj@njit.edu, haimin.wang@njit.edu}
\affiliation{Department of Computer Science, New Jersey Institute of Technology, University Heights, Newark, NJ 07102-1982, USA}

\author{Haimin Wang}
\affiliation{Institute for Space Weather Sciences, New Jersey Institute of Technology, University Heights, Newark, NJ 07102-1982, USA;
hj78@njit.edu, wangj@njit.edu, haimin.wang@njit.edu}
\affiliation{Center for Solar-Terrestrial Research, New Jersey Institute of Technology, University Heights, Newark, NJ 07102-1982, USA}
\affiliation{Big Bear Solar Observatory, New Jersey Institute of Technology, 40386 North Shore Lane, Big Bear City, CA 92314-9672, USA}

\begin{abstract}
Obtaining high-quality magnetic and velocity fields through Stokes inversion 
is crucial in solar physics.
In this paper, we present a new deep learning method, named Stacked Deep Neural Networks (SDNN), for inferring line-of-sight (LOS) velocities and Doppler widths from Stokes profiles collected by the Near InfraRed Imaging Spectropolarimeter (NIRIS) on the 1.6 m Goode Solar Telescope (GST) at the Big Bear Solar Observatory (BBSO).
The training data of SDNN is prepared by a Milne–Eddington (ME) inversion code
used by BBSO.
We quantitatively assess SDNN, comparing its inversion results
with those obtained by the ME inversion code and related machine learning (ML)
algorithms such as multiple support vector regression,
multilayer perceptrons and
a pixel-level convolutional neural network.
Major findings from our experimental study are summarized as follows.
First, the SDNN-inferred LOS velocities are highly correlated to the ME-calculated ones
with the Pearson product-moment correlation coefficient being close to 0.9 on average.
Second, SDNN is faster, while producing smoother and cleaner LOS
velocity and Doppler width maps, than the ME inversion code.
Third, the maps produced by SDNN are closer to ME's maps than those from
the related ML algorithms, 
demonstrating the better learning capability of SDNN than the ML algorithms.
Finally, comparison between the inversion results of ME and SDNN based on GST/NIRIS and
those from the Helioseismic and Magnetic Imager 
on board the Solar Dynamics Observatory
in flare-prolific active region NOAA 12673
is presented. 
We also discuss extensions of SDNN
for inferring vector magnetic fields 
with empirical evaluation.
\end{abstract}


\section{Introduction} 
\label{sec:intro}

Obtaining high-quality magnetic and velocity fields through Stokes inversion is crucial  
in solar physics \citep{1998A&AS..127..607N, 2010ASSP...19..517M, 2015ApJ...798..135B}.
Stokes inversion attempts to infer the physical conditions of the solar atmosphere
based on the interpretation of observed Stokes profiles \citep{2008ApJ...683..542A, 2016JGRA..121.5025B}. 
Estimates of the physical magnitudes governing the state of the solar atmosphere
can be obtained through the various inversion methods
that try to achieve the best fit to the observed Stokes profiles \citep{2014RAA....14.1469T}.
An excellent review of the status of inversions can be found in
\citet{2016LRSP...13....4D},
in which the Milne–Eddington (ME) approximation is commonly used.
The ME is an approach to the radiative transfer equation (RTE) which states that when all the atmospheric quantities are constant with depth except for the source function that varies linearly,
the RTE can be solved analytically \citep{1956PASJ....8..108U,1962IzKry..28..259R, 1963IzKry..30..267R,1977SoPh...55...47A, 1982MmSAI..53..841L, 1982SoPh...78..355L, 1983SoPh...85....3L, 2004ASSL..307.....L}.
Implementations based on the ME approach include
Helix+ \citep{2004A&A...414.1109L},  MILOS \citep{2007ApJ...670L..61O}, MERLIN \citep{2007MmSAI..78..148L} and VFISV \citep{2007SoPh..240..177B, 2011SoPh..273..267B}. 
On the other hand, with the availability of high performance computing,
algorithms based on 
the local thermodynamic equilibrium (LTE) and non-LTE conditions,
which can solve the full radiative transfer equation,
also become popular.
Examples of such algorithms include 
SIR \citep{1992ApJ...398..375R}, SPINOR \citep{2000A&A...358.1109F}, 
and NICOLE \citep{2015A&A...577A...7S}. 

Since Stokes inversion is a time-consuming task,
there have been efforts of employing machine learning (ML) to
accomplish this task.
After an ML model is trained, one can use the trained model
to perform Stokes inversion through making predictions,
which reduces the inversion time significantly \citep{Liu_2020}.
For example, \citet{2015SoPh..290.2693T} developed
a multiple support vector regression (MSVR) method
for Stokes inversion.
The MSVR method took as input
Stokes \textit{I}, \textit{Q}, \textit{U}, \textit{V} profiles
scaled at six wavelengths from the Helioseismic and Magnetic Imager (HMI) on board
the Solar Dynamics Observatory (SDO)
and produced as output
atmospheric parameters including
the magnetic field strength and inclination angle.
\citet{Carroll_2008} designed 
three multilayer perceptrons (MLPs),
which took as input
Stokes \textit{I}, \textit{V} profiles resulted from 
 magnetohydrodynamic (MHD) simulations
 and produced as output
 the line-of-sight (LOS) velocity, 
 magnetic field and temperature, respectively.

In recent years, more powerful deep learning approaches have been developed 
to solve the Stokes inversion problem. 
For example, \citet{Asensio_2019} presented two convolutional neural networks \citep[CNNs;][]{LeCun2015} 
to perform Stokes inversion on synthetic two-dimensional (2D) maps of Stokes profiles
where the authors exploited the 2D spatial coherence of the field of view 
to produce a three-dimensional (3D) cube of thermodynamic and magnetic properties.
\citet{Liu_2020} designed 
a pixel-level CNN, referred to as PCNN, 
to perform Stokes inversion 
on the Near InfraRed Imaging Spectropolarimeter (NIRIS) data \citep{2012ASPC..463..291C}
from the 1.6 m Goode Solar Telescope (GST) at the Big Bear Solar Observatory \citep[BBSO;][]{2010AN....331..636C, 2010ApJ...714L..31G, 2012SPIE.8444E..03G, 2014SPIE.9147E..5DV}.
The PCNN took as input Stokes 
\textit{Q}, \textit{U}, \textit{V} profiles from GST/NIRIS and
produced as output 
three components of the magnetic field vector 
(i.e., the magnetic field strength, inclination angle and azimuth angle).
Separately, \citet{2020A&A...644A.129M}
developed a similar PCNN,
which took as input 
Stokes \textit{I}, \textit{V} profiles resulted from MHD simulations and
produced as output
the magnetic field strength,
LOS velocity,
microturbulent velocity,
temperature
and magnetic field inclination.
Later, the authors extended their approach to 
Stokes \textit{I}, \textit{Q}, \textit{U}, \textit{V} profiles and
used the PCNN-inverted results to accelerate the SIR inversion code \citep{2021A&A...651A..31G}.

In this paper, we present a new deep learning method, named
Stacked Deep Neural Networks (SDNN), for 
inverting Stokes profiles of GST/NIRIS.
SDNN aims to infer two kinematic parameters, 
namely the LOS velocity and Doppler width.
In addition, SDNN can be extended to infer 
the magnetic field strength, inclination angle and azimuth angle.
These five atmospheric parameters including
the LOS velocity, Doppler width,
magnetic field strength, inclination angle and azimuth angle
are mostly used by researchers to understand the evolution of 
physical properties of the solar atmosphere \citep{2021RSPTA.37900182K}.
It should be pointed out that although both SDNN and
the PCNN developed by \citet{Liu_2020}
perform Stokes inversion on the data from GST/NIRIS,
the two tools differ in three ways.
First, Liu et al.'s PCNN focuses on predicting the magnetic field strength, inclination angle and azimuth angle. 
In contrast, SDNN is designed to infer the LOS velocity and Doppler width 
in addition to the vector magnetic field.
Second, when applying Liu et al.'s PCNN to infer the LOS velocity,
it fails to infer the granular patterns in the LOS velocity image of sunspot data.
In contrast, SDNN can infer all the
convective granulation structures in the LOS velocity image \citep{Ortiz_2014}. 
Third, the architecture of SDNN, which is better suited for Stokes inversion,
is totally different from that of Liu et al.'s PCNN.
As demonstrated in our experimental study, 
SDNN outperforms Liu et al.'s PCNN on several different datasets.

The rest of this paper is organized as follows.
Section \ref{sec:observational data} describes solar observations and GST/NIRIS data used in this study.
Section \ref{sec:method} presents details of our SDNN model. 
Section \ref{sec:experiment} reports experimental results
obtained by using SDNN to infer LOS velocities and Doppler widths
from Stokes profiles of GST/NIRIS.
We also discuss results obtained by extending SDNN to infer
magnetic field strengths, inclination and azimuth angles
from the Stokes profiles of GST/NIRIS.
Section \ref{sec:conclusion} presents a discussion and concludes the paper.

\section{Observations and Data Preparation}
\label{sec:observational data}

GST/NIRIS is a Fabry-P$\acute{e}$rot based imaging system, 
which provides high-resolution Stokes parameters \textit{I}, \textit{Q}, \textit{U} and \textit{V} 
of the \ion{Fe}{1} 1560 nm line within a $\pm 0.25$ nm spectral window \citep{2012ASPC..463..291C}. 
A typical field-of-view is about 85\arcsec \hspace*{+0.01cm}
with an image scale of 0.\arcsec083 pixel$^{-1}$ 
\citep{2015NatCo...6.7008W, Xu_2016, Liu_2018, 2018NatCo...9...46X}.
The data used in this study were obtained from three active regions (ARs), namely AR 12371, AR 12665, and AR 12673. 
Our first dataset, denoted D1, is a $990 \times 950$ image from AR 12371 collected at 17:33:00 UT on 2015 June 22.
The second dataset, denoted D2, is a $720 \times 720$ image 
from the same AR 12371 but
collected approximately three days later at 16:55:13 UT on 2015 June 25.
The third dataset, denoted D3, is a $720 \times 720$ image 
from AR 12665
collected at 16:20:12 UT on 2017 July 13.
The fourth dataset, denoted D4,
is a $720 \times 720$ image 
from AR 12673
collected at 19:17:53 UT on 2017 September 6. 

Since D1 has the most pixels with 
the largest range of Stokes component values
among the four datasets,
we use D1 as the training set.
This dataset has $990 \times 950$ = 940,500 pixels. 
Each pixel is treated as a training data sample containing Stokes component values 
and labeled by the LOS velocity and Doppler width calculated by 
an ME inversion code.
The code is specifically designed for GST/NIRIS \citep{2009ASPC..415..101C, 2017SPD....4811504A, 2019ASPC..526..317A}. 
Its early version was applied to the Hinode/SP data \citep{2009ASPC..415..101C}. 
The code, written with the IDL language, is based on the formulae given in \citet{1992soti.book...71L}. 
After careful elimination of the cross talk among Stokes profiles \citep{2019ASPC..526..317A}, we perform Stokes inversion under the ME atmospheric approximation (with initial parameters pre-calculated to resemble the observed Stokes profiles), assuming that the magnetic field and velocity are constant with height through the solar atmosphere \citep{2014A&A...572A..54B}.
We set the filling factor/stray light fraction parameter to unity, because magnetic structures are believed to be fully resolved in these data \citep{2019ApJ...873...75S}.
The code outputs 
9
parameters including 
the Doppler shift, Doppler width, magnetic field strength, inclination angle, azimuth angle, and so on.
We assign Stokes \textit{I} profiles less than a half weight than 
Stokes \textit{Q}, \textit{U}, \textit{V} profiles 
before performing the inversion fitting process 
as the Stokes \textit{I} profiles intrinsically have more noise factors 
than the other profiles. 
The fitting process may return error values for the inverted  
9
parameters.
Most of the fitting errors come from the initial guess of the longitudinal field strength calculated by the center-of-gravity method \citep{2018tess.conf30818A}.

Thus, as in \citet{Liu_2020}, we use the output of the ME inversion code
as the training labels.
Notice that the number of spectral points scanned by GST/NIRIS is usually 60, 
but varied in some particular days. 
For instance, there were 56 spectral points 
for AR 12673
scanned on 2017 September 6.
For consistency reasons, zeros are added so that the number of spectral points is unified 
and fixed at 60. 
There are four Stokes components \textit{I}, \textit{Q}, \textit{U}, \textit{V} at each spectral position, 
so the length of each training data sample, corresponding to each pixel in D1, 
is $60 \times 4$.
There are two labels, namely the LOS velocity and Doppler width
produced by the ME code,
associated with the pixel.
Therefore, the total length of the training data sample
fed to our SDNN model
is $60 \times 4 + 2$ = 242.

The remaining three datasets, D2, D3, D4, are used as test sets.
The training set and test tests are disjoint.
Hence, our SDNN model is tested on data that the model has never seen during training.
Each test data sample in the test sets (D2, D3, D4) has the same format 
as the training data samples in D1 except that 
the test data sample does not have the two labels.
Therefore, the length of the test data sample is $60 \times 4$ = 240.
Because the values of the Stokes components vary,
to facilitate machine learning \citep{Liu_2020},
we normalize the Stokes \textit{Q}, \textit{U}, \textit{V} profiles 
by dividing them by 1000,
as most of the Stokes measurements have values between $-1000$ and $+1000$.
In addition, we normalize the Stokes \textit{I} profile by dividing the measurements by 10000, 
since the mean of the measurements is around 10000. 
The Doppler shift values, which are obtained directly from the ME code and
can be converted to the LOS velocities as we explain later, range from $-0.5$ to $0.5$.
We normalize the Doppler shift values by adding $0.5$ to all the values, so that they range from 0 to 1.
The Doppler width values calculated by the ME code
already range from 0 to 1, and hence no normalization is done and
the Doppler width values are used directly for model training.

\section{Methodology} 
\label{sec:method}

\begin{figure}
	\epsscale{1.15}
	\plotone{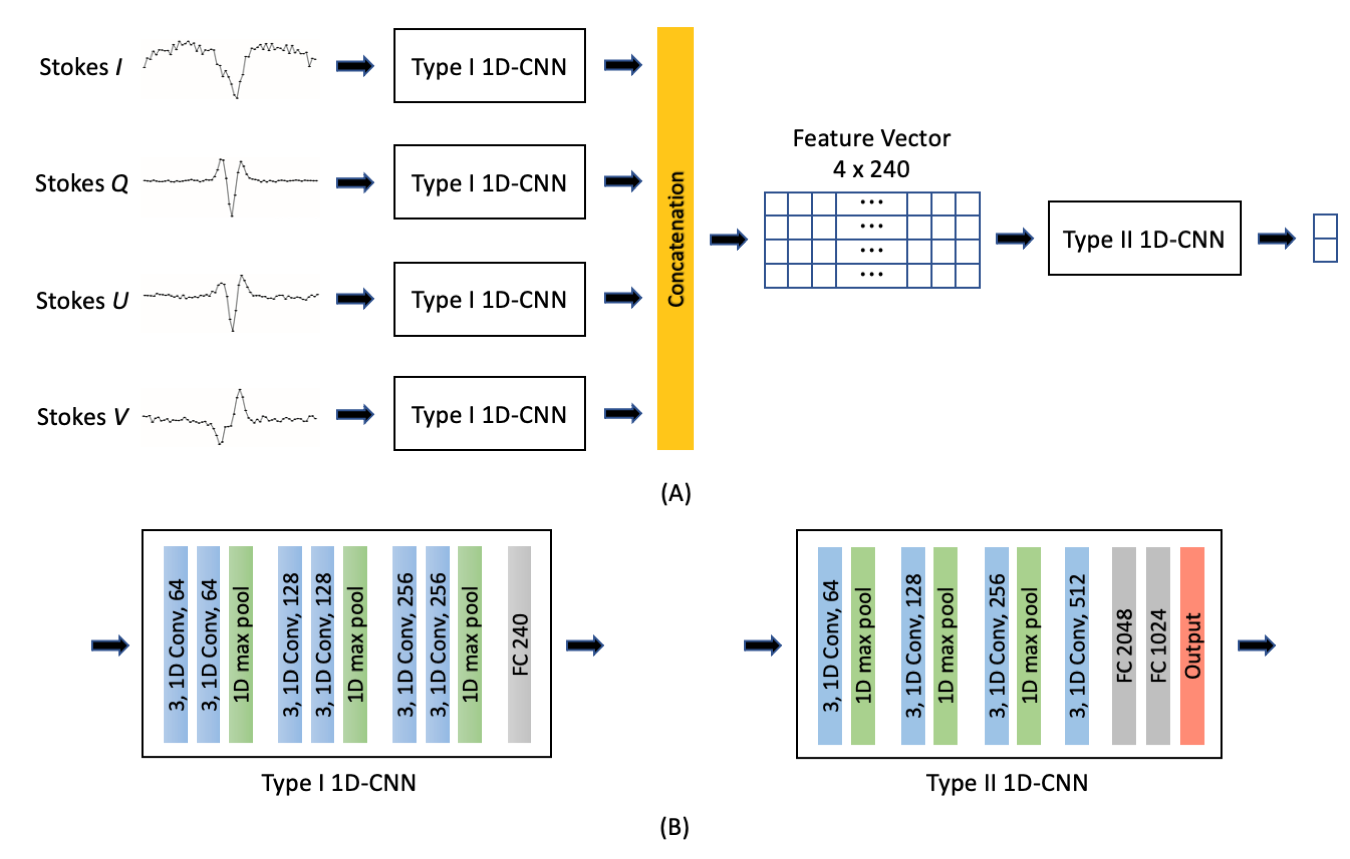}
	\caption{(A) Architecture of our SDNN model. 
	The model contains two types of 1D convolution neural network (1D-CNN),
	named Type I 1D-CNN and Type II 1D-CNN respectively.
	There are four Type I 1D-CNNs, which are 
	stacked on top of a Type II 1D-CNN;
	hence the name SDNN (Stacked Deep Neural Networks) is used.
	The input of the SDNN is a sequence of four Stokes 
\textit{I}, \textit{Q}, \textit{U} and \textit{V} components where
each component has 60 wavelength sampling points and the four Stokes components
correspond to a pixel.
Each Type I 1D-CNN takes as input a Stokes component,
and encodes and produces as output a 240-dimensional feature vector.
There are four Stokes components so the four Type I 1D-CNNs output
four 240-dimensional feature vectors,
which are concatenated to form a $4 \times 240$ feature vector.
The Type II 1D-CNN takes as input the $4 \times 240$ feature vector and
produces as output
two numbers representing
the estimated 
Doppler shift and Doppler width of the input pixel.
(B) Configuration details of the Type I 1D-CNN and
Type II 1D-CNN used in our SDNN model.
}
\label{fig:model}
\end{figure}

Figure \ref{fig:model} illustrates the architecture of our deep learning model (SDNN) 
used to infer 
LOS velocities and Doppler widths 
from Stokes profiles of GST/NIRIS.
This model contains two types of 1D convolution neural network 
\citep[1D-CNN;][]{8682194}. 
Type I 1D-CNN contains 3 convolution blocks followed by a fully connected (FC) layer with 240 neurons. 
Each convolution block contains two 1D convolution layers with kernels of size 3, activated 
by a ReLU (rectified linear unit) function, followed by a 1D max pooling layer with a kernel of size 2. 
Each 1D convolution layer in the first (second and third, respectively) convolution block
has 64 (128 and 256, respectively) kernels.

Type II 1D-CNN contains four convolution blocks followed by two fully connected layers
activated by ReLU and 
having 2048 and 1024 neurons, respectively.
Each of the first three convolution blocks contains a 1D convolution layer with a kernel of size 3 activated by ReLU, followed by a 1D max pooling layer with a kernel of size 2.
The fourth convolution block only contains a 1D convolution layer with a kernel of size 3 activated by ReLU.
The 1D convolution layer
in the first (second, third and fourth, respectively) convolution block has
64 (128, 256 and 512, respectively) kernels. 
The output layer has 
2 neurons, activated by the linear function, $f(x) = ax$ \citep{Goodfellow-et-al-2016}, which is suitable for predicting continuous numerical values and performs better than 
the Tanh function used in \citet{Liu_2020}.

The input of our SDNN model is a sequence of four Stokes 
\textit{I}, \textit{Q}, \textit{U} and \textit{V} components where
each component has 60 wavelength sampling points and
the four Stokes components
correspond to a pixel.
There are four Type I 1D-CNNs
where each Type I 1D-CNN
takes as input a Stokes component,
and encodes and produces as output a 240-dimensional feature vector.
There are four Stokes components so the four Type I 1D-CNNs output four 240-dimensional feature vectors,
which are concatenated to form a $4 \times 240$ feature vector.
The Type II 1D-CNN takes as input the $4 \times 240$ feature vector and
produces as output 
two numbers representing the estimated 
Doppler shift and Doppler width
of the input pixel.

The regression loss function used by our SDNN model is 
the L1 loss function \citep{Goodfellow-et-al-2016}, defined as follows:
\begin{equation}
\mbox{L1 loss}
=\frac{1}{N} \sum_{i=1}^{N} (\left|y_{i}^{ds}-\hat{y}_{i}^{ds}\right|
+ \left|y_{i}^{dw}-\hat{y}_{i}^{dw}\right|)
 \label{loss}
\end{equation}
where $N$ is the total number of pixels/data samples in a test set,
$y_{i}^{ds}$ ($y_{i}^{dw}$, respectively) is 
the Doppler shift (Doppler width, respectively)
of the $i$th pixel calculated by the ME inversion code
described in Section \ref{sec:observational data},
$\hat{y}_{i}^{ds}$ ($\hat{y}_{i}^{dw}$, respectively)
is the Doppler shift (Doppler width, respectively)
of the $i$th pixel predicted by our SDNN model.
We use the L1 loss function here
because it is more robust to outliers, 
hence making the model more tolerant to noise in the training data
\citep{10.2307/24869236}.

We train the SDNN model using the Adam optimizer. 
The batch size is set to 1024, and the number of epochs is set to 40. 
During testing, the trained model takes as input 
the GST/NIRIS Stokes \textit{I}, \textit{Q}, \textit{U}, and \textit{V} profiles 
of each pixel in a test set,
and predicts as output the Doppler shift and
Doppler width of the pixel. 
The predicted values are denormalized so that
they fall in the original (correct) range.
We then convert the Doppler shift,
denoted $\Delta_{\lambda}$,
to the LOS velocity, denoted $v_{LOS}$, as follows:
\begin{equation}
v_{LOS} = \frac{C \times \Delta_{\lambda}}{ \lambda} 
\end{equation}
where 
$\lambda$ is the GST/NIRIS magnetogram wavelength, 
which is set to 1.56 $um$, and $C$ is the speed of light. 
The unit of $v_{LOS}$ is km/s.  

\section{Results} 
\label{sec:experiment}

\subsection{Performance Metrics}

For each test data sample, which corresponds to each pixel in a test set,
we can use the proposed SDNN model to predict or infer its 
LOS velocity and Doppler width.
In addition, we can also use the ME inversion code
described in Section \ref{sec:observational data}
to calculate its LOS velocity and Doppler width.
We adopted four metrics to
evaluate the performance of our SDNN model
and compare it with related machine learning algorithms.
Specifically, we considered two quantities: LOS velocity and Doppler width.
For each quantity,
we compared its ME-calculated values with our SDNN-inferred
values and computed the four performance metrics.

The first performance metric is the mean absolute error \citep[MAE;][]{MAE}.
MAE quantitatively assesses the difference between the
ME-calculated and SDNN-inferred values for the test set (test image).
The smaller the MAE is, the better performance an algorithm has.
The second performance metric is the percent agreement \citep[PA;][]{PA}.
Let $y_{i}$ and $\hat{y}_{i}$ denote the ME-calculated and
SDNN-inferred value respectively for the $i$th pixel in the test image.
The $i$th pixel is an agreement pixel if 
$|y_{i}-\hat{y}_{i}|$ is smaller than a threshold. 
(The default threshold is set to  1 km/s for the LOS velocity and
0.1  \AA \hspace*{+0.01cm} for the Doppler width.)
PA equals the number of agreement pixels
divided by the total number of pixels in the test image
multiplied by 100\%.
Thus PA quantitatively assesses the similarity between the ME-calculated
and SDNN-inferred values for the test image. 
The closer to 100\% the PA is,
the better performance an algorithm has.
The third performance metric is R-squared \citep{R2}. R-squared is a statistical measure that determines the strength of the relationship between the ME-calculated and SDNN-inferred values for the test image. The values of R-squared range from $-\infty$ to $+1$. The closer to $+1$ R-squared is, the stronger the relationship between the ME-calculated and SDNN-inferred values is. 
The fourth performance metric is the Pearson product-moment correlation coefficient \citep[PPMCC;][]{PPMCC}. PPMCC measures the linear correlation between the ME-calculated and SDNN-inferred values for the test image. The values of PPMCC range from $-1$ to $+1$. A PPMCC of $+1$ indicates a perfect positive correlation while a PPMCC of $-1$ indicates a perfect negative correlation.

\subsection{Impact of Training Data on the Performance of the SDNN Method}

Table \ref{table: metrics on three ARs} presents 
experimental results of using D1 as the training set
to train SDNN and using
D2, D3 and D4 as test sets to test SDNN as described in 
Section \ref{sec:observational data}.
SDNN works well on D2 and D3.
However, the performance of SDNN degrades on D4 which contains pixels
from AR 12673.
We note that AR 12673 is the most flare-productive active region in solar cycle 24
\citep{Wang_2018_inversion}.
The training set D1 only contains pixels from 
a normal active region (AR 12371);
the SDNN model trained by D1 does not have sufficient knowledge about 
the extremely flare-productive active region AR 12673.
To assess and quantify the impact of training data on the performance of SDNN,
we additionally selected a $720 \times 720$ image with 518,400 pixels (data samples)
from AR 12673 collected at 16:18:41 UT on 2017 September 6.
We referred to this additional dataset as D5.
Thus, the image of D5 was taken approximately 3 hours before the image of D4.
We then combined the 940,500 pixels in D1 and the 518,400 pixels in D5
to get a new training set, denoted D1 $\cup$ D5.
D1 $\cup$ D5 contains 1,458,900 training data samples (pixels) in total.
Results of using D1 $\cup$ D5 as the training set
to train SDNN and using
D2, D3 and D4 as test sets to test SDNN
are also presented in 
Table \ref{table: metrics on three ARs}.
Notice that, due to the time difference, 
D1 $\cup$ D5 and D4 are disjoint even though
the data samples in D5 and D4 are from the same AR 12673.

We see from Table \ref{table: metrics on three ARs} 
that the SDNN model trained by D1 $\cup$ D5
outperforms the SDNN model trained by D1 
when the two models are tested on D4.
This happens because the SDNN model trained by D1 $\cup$ D5
acquires more knowledge concerning D4 
than the SDNN model trained by D1 
due to the fact that D4 and D5 are from the same AR 12673 as indicated above.
On the other hand, the two models have similar performance 
when tested on D2 and D3.
Notice that D1, D2 and D3 
are from normal active regions (AR 12371 and AR 12665) while
D4 contains special pixels 
from an extremely flare-productive active region (AR 12673).
The special pixels (data samples) in D4 do not occur in D1, D2, D3.
Thus, the model trained by D1 lacks knowledge of 
the special pixels,
and hence does not work well on D4.
On the other hand, like D4, D5 also contains special pixels 
from the extremely flare-productive AR 12673.
As a consequence, the model trained by D1 $\cup$ D5
performs well on D4.
These results indicate that 
when dealing with normal ARs,
the training data samples in D1 are sufficient to produce good results.
On the other hand, when dealing with special ARs such as AR 12673,
the training set must be expanded to include data samples 
from the special ARs
so that our SDNN model can acquire sufficient knowledge about the special ARs
to produce good results.
In view of the experimental results, 
we used D1 $\cup$ D5 as the training set
in subsequent experiments.

\begin{table}[t]
	\caption{Performance Metric Values of SDNN Based on Two Training Sets and Test Images from Three ARs}
	\label{table: metrics on three ARs}
	\centering
	\begin{tabular*}{0.9999\textwidth}{@{\extracolsep{\fill}} cccccc}
		\toprule
		\toprule
        \multirow{2}{*}{} & \multirow{2}{*}{} & \multicolumn{2}{c}{LOS Velocity} & \multicolumn{2}{c}{Doppler Width} \\
        \hspace{0.85cm}Test Image & Metric & D1 & D1 $\cup$ D5 & D1 & D1 $\cup$ D5 \\ \midrule
		\multirow{4}{*}{\begin{tabular}[c]{@{}c@{}}D2\\2015 June 25 \\ 16:55:13 UT \\ (AR 12371)\end{tabular}}
		& MAE & 0.251 & 0.260 & 0.048 & 0.047  \\
		& PA & 97.8$\%$ & 97.5$\%$ & 89.1$\%$ & 90.3$\%$  \\  
		& R-squared & 0.796 & 0.786 & 0.317 & 0.458  \\
		& PPMCC & 0.915 & 0.912 & 0.780 & 0.787  \\
		\midrule
		\multirow{4}{*}{\begin{tabular}[c]{@{}c@{}}D3\\2017 July 13 \\ 16:20:12 UT \\ (AR 12665)\end{tabular}} 
		& MAE & 0.289 & 0.302 & 0.039 & 0.040  \\
		& PA & 97.6$\%$ & 97.0$\%$ & 93.2$\%$ & 93.2$\%$ \\
		& R-squared & 0.768 & 0.728 & 0.224 & 0.278 \\ 
		& PPMCC & 0.912 & 0.896 & 0.716 & 0.723\\
		\midrule 
		\multirow{4}{*}{\begin{tabular}[c]{@{}c@{}}D4\\2017 September 6 \\ 19:17:53 UT \\ (AR 12673)\end{tabular}} 
		& MAE & 0.952 & 0.316 & 0.064 & 0.034  \\
		& PA & 51.0$\%$ & 94.5$\%$ & 82.3$\%$ & 95.3$\%$ \\ 
		& R-squared & -0.402 & 0.749 & -1.334 & 0.208  \\ 
		& PPMCC & 0.804 & 0.883 & 0.516 & 0.711  \\ 
		\bottomrule 
	\end{tabular*}
\end{table}

\subsection{Comparison between the SDNN and ME Methods}

\begin{figure}
	\epsscale{1.15}
	\plotone{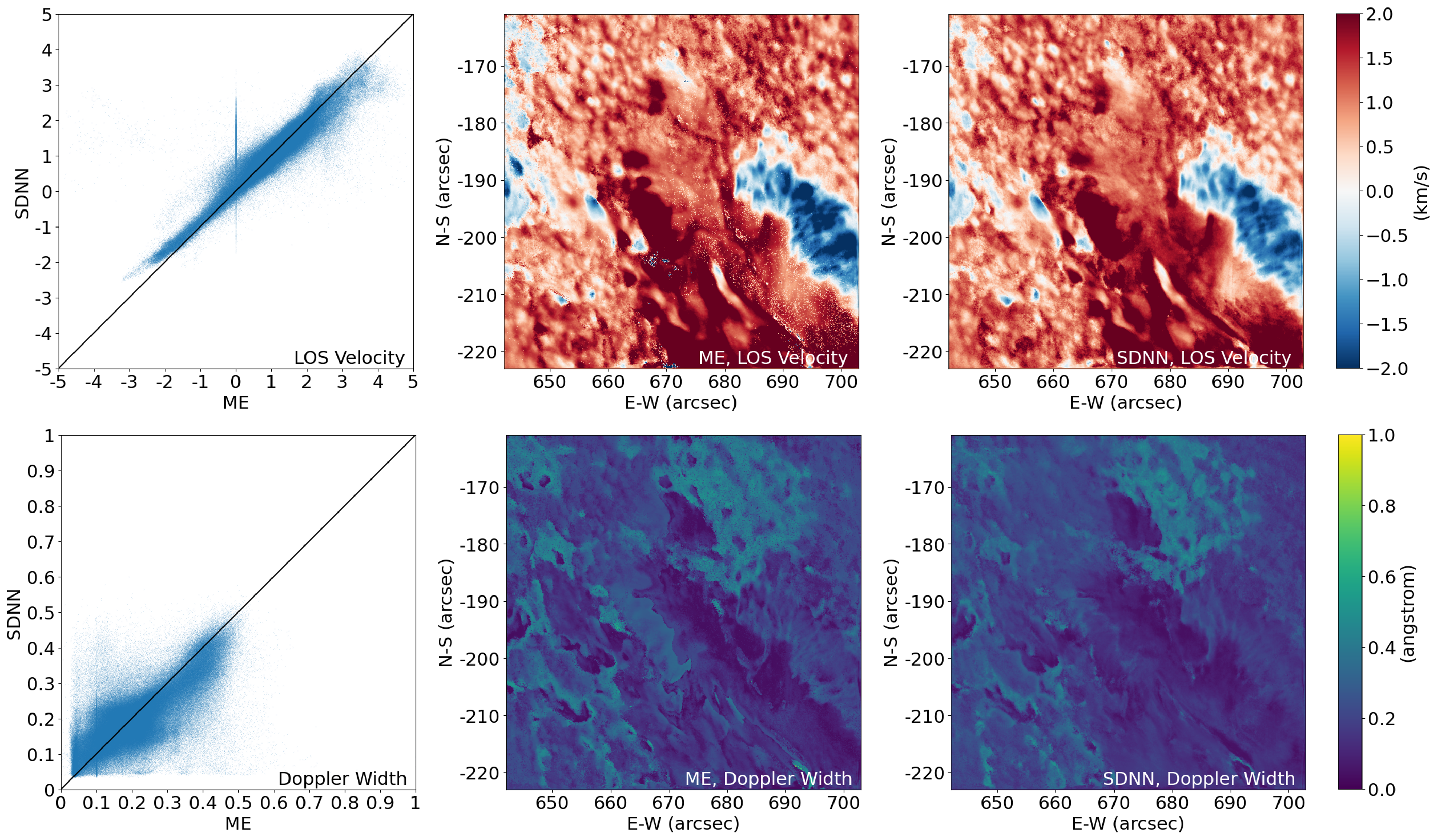}
	\caption{Comparison between the ME inversion code and SDNN for producing the LOS velocity maps (top panels) and Doppler width maps (bottom panels) based on the test image/dataset D2 from AR 12371 collected on 2015 June 25 16:55:13 UT, where training data were taken from D1 $\cup$ D5. The first column shows scatter plots where the X-axis and Y-axis represent the values obtained by the ME code and SDNN respectively.
	Pixels on the diagonal line in a scatter plot 
    have identical ME-calculated and SDNN-inferred values.
	The second and third columns show the maps produced by the ME code and SDNN respectively.} 
	\label{fig:results_150625}
\end{figure}

\begin{figure}
	\epsscale{1.15}
	\plotone{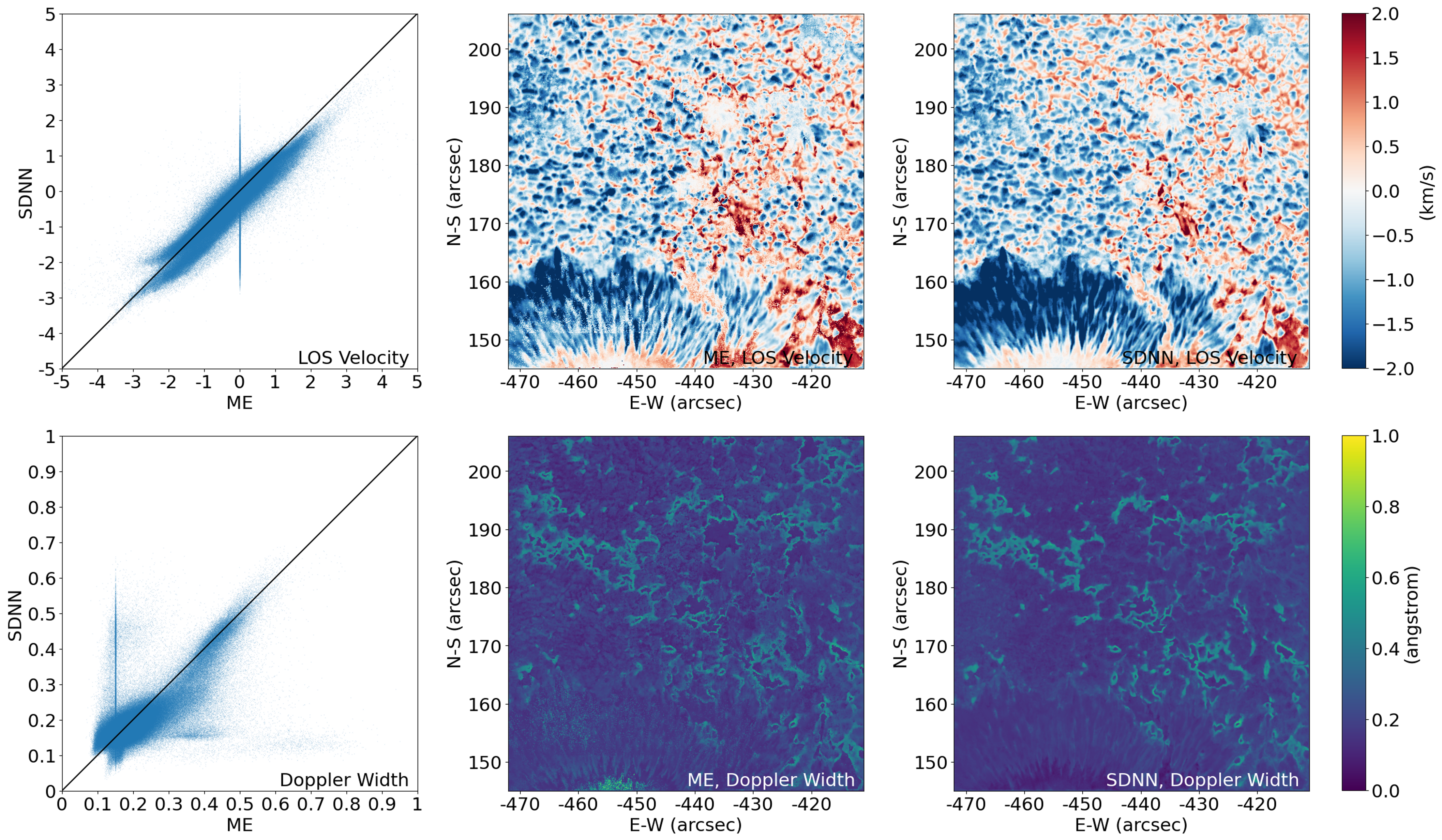}
	\caption{Comparison between the ME inversion code and SDNN for producing the LOS velocity maps (top panels) and Doppler width maps (bottom panels) based on the test image/dataset D3 from AR 12665 collected on 2017 July 13 16:20:12 UT, where training data were taken from D1 $\cup$ D5. The first column shows scatter plots where the X-axis and Y-axis represent the values obtained by the ME code and SDNN respectively.
	Pixels on the diagonal line in a scatter plot 
    have identical ME-calculated and SDNN-inferred values.
	The second and third columns show the maps produced by the ME code and SDNN respectively.} 
	\label{fig:results_170713}
\end{figure}

\begin{figure}
	\epsscale{1.15}
	\plotone{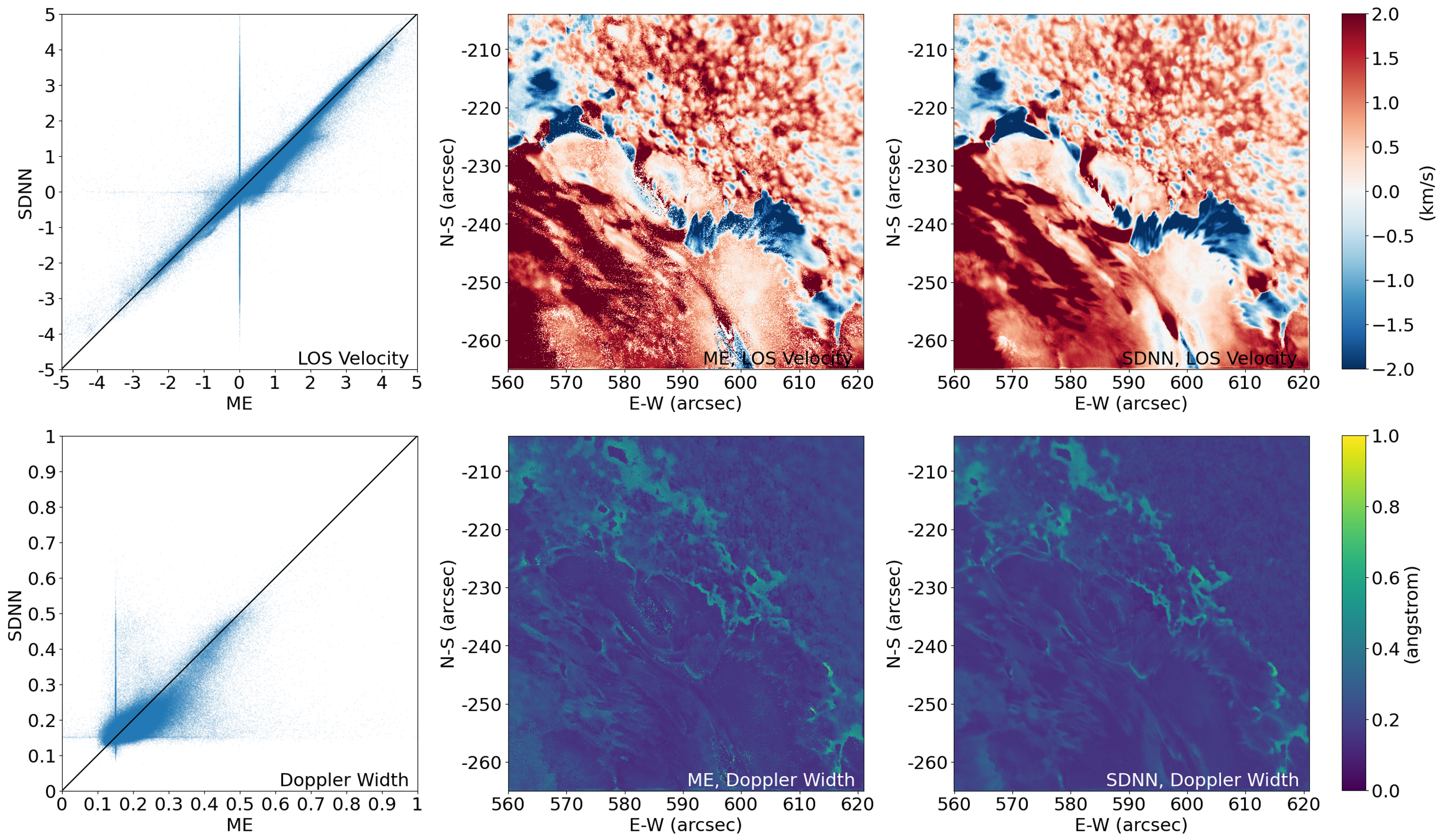}
	\caption{Comparison between the ME inversion code and SDNN for producing the LOS velocity maps (top panels) and Doppler width maps (bottom panels) based on the test 
	image/dataset D4 from AR 12673 collected on 2017 September 6 19:17:53 UT, where training data were taken from D1 $\cup$ D5. The first column shows scatter plots
	where the X-axis and Y-axis represent the values obtained by the ME code and SDNN respectively.
	Pixels on the diagonal line in a scatter plot 
    have identical ME-calculated and SDNN-inferred values.
	The second and third columns show the maps produced by the ME code and SDNN respectively.} 
	\label{fig:results_170906}
\end{figure}

Here we compare the LOS velocity maps and Doppler width maps 
produced by the SDNN and ME methods. 
Figure \ref{fig:results_150625} (Figure \ref{fig:results_170713}, 
Figure \ref{fig:results_170906}, respectively) 
presents the LOS velocity maps (top panels) and Doppler width maps (bottom panels)
produced by the two methods 
based on the test set/image D2 (D3, D4, respectively) with data samples/pixels from
AR 12371 (AR 12665, AR 12673, respectively)
collected on 2015 June 25 16:55:13 UT (2017 July 13 16:20:12 UT, 2017 September 6 19:17:53 UT, respectively). 
The first columns in the figures show scatter plots, 
the second columns show the maps produced by the ME inversion code
described in Section \ref{sec:observational data},
and the third columns show the maps produced by the SDNN model.
Pixels on the diagonal line in a scatter plot 
have identical ME-calculated and SDNN-inferred values.
We see from Figures \ref{fig:results_150625}, \ref{fig:results_170713} and \ref{fig:results_170906}
that the maps 
produced by the ME code and SDNN model are highly correlated.
The high correlation can be seen particularly from the scatter plots
of the LOS velocity maps and the corresponding PPMCC
values of $\sim$0.9 as shown in 
Table \ref{table: metrics on three ARs}.
Since the training labels for the SDNN model are produced by the ME inversion code
described in Section \ref{sec:observational data},
these results demonstrate the good learning capability of SDNN.
The Doppler width inferences, as seen from the figures, show larger scatter. This happens because of the degeneracy of the Doppler width with the rest of the ME thermodynamic parameters as described in \citet{2007A&A...462.1137O}.
Notice also that 
the maps produced by SDNN are smoother and cleaner than those
produced by the ME code.
We see many salt-and-pepper noise pixels in the maps of the ME code.
The many noise pixels from the ME code are also reflected in the
scatter plots in Figures \ref{fig:results_150625}, \ref{fig:results_170713} and \ref{fig:results_170906}.
For example, refer to the scatter plot of the LOS velocity maps
in Figure \ref{fig:results_170713} where
there is a vertical line on which pixels have a LOS velocity of zero.
Many of the pixels on the vertical line are noisy ones in the LOS velocity map 
produced by the ME code.

\begin{figure}
\epsscale{1.15}
\plotone{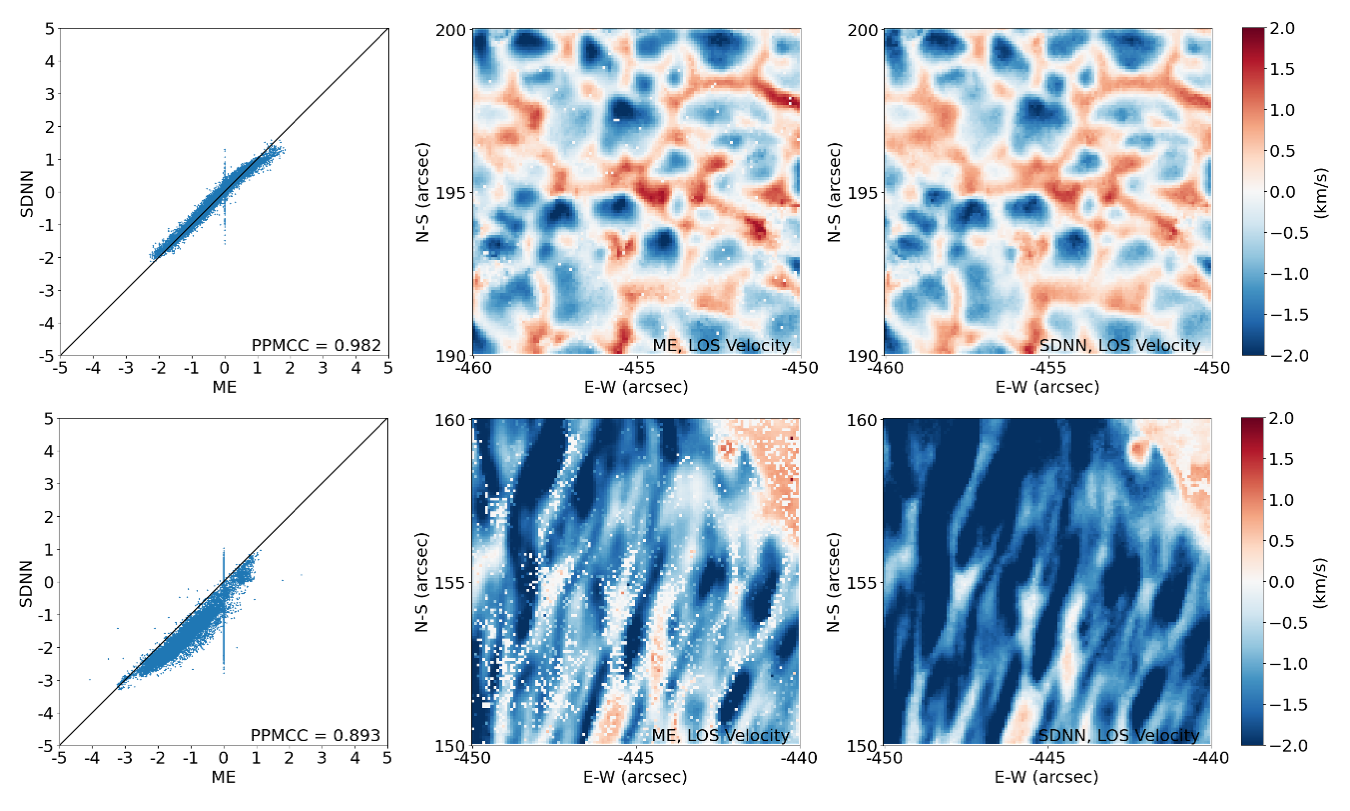}
\caption{Comparison between the ME inversion code and SDNN for producing the local LOS 
velocity maps containing granular patterns (top panels) and a portion of a 
sunspot penumbra (bottom panels)
based on the test image/dataset D3 from AR 12665 collected on 2017 July 13 16:20:12 UT, where training data were taken from D1 $\cup$ D5.
The first column shows scatter plots
where the X-axis and Y-axis represent the values obtained by the ME code and SDNN respectively.
Pixels on the diagonal line in a scatter plot 
have identical ME-calculated and SDNN-inferred values.
The second and third columns show the local LOS velocity maps produced by the ME code and SDNN respectively.} 
\label{fig:granular_penumbral}
\end{figure}

It is worth noting that the LOS velocity maps in Figure \ref{fig:results_170713}
contain granular patterns
and a portion of a sunspot penumbra.
For example, there are granular patterns located in the region 
whose E-W coordinates are between $-460$\arcsec \hspace*{+0.01in} and $-450$\arcsec \hspace*{+0.01in} and N-S coordinates are between 
190\arcsec \hspace*{+0.01in} and 200\arcsec.
A partial sunspot penumbra is located in the region
whose E-W coordinates are between 
$-450$\arcsec \hspace*{+0.01in} and $-440$\arcsec
\hspace*{+0.01in} and N-S coordinates are between 
150\arcsec \hspace*{+0.01in} and 160\arcsec.
Figure \ref{fig:granular_penumbral} presents an enlarged view of
these two regions.
Both ME and SDNN produce the granular patterns with a PPMCC of 0.982
and the partial sunspot penumbra with a PPMCC of 0.893.
The similarity between the local maps produced by the two methods
is also reflected in the scatter plots in Figure \ref{fig:granular_penumbral}.
In addition, we see from Figure \ref{fig:granular_penumbral} that
the local maps produced by the SDNN model are smoother and cleaner than
those produced by the ME code.
There are many salt-and-pepper noise pixels in the maps of the ME code, 
particularly in the region of the partial sunspot penumbra.

\subsection{Comparison with Related Machine Learning Methods} 

Here we compare SDNN with three related machine learning (ML) methods
including 
MSVR (multiple support vector regression) \citep{DBLP:conf/kes/ReesGAG04},
MLP (multilayer perceptrons)  \citep{Carroll_2008} and 
PCNN (pixel-level convolutional neural network) \citep{Liu_2020},
all of which have previously been used in Stokes inversion
as surveyed in Section \ref{sec:intro}.
The MSVR method uses the radial basis function
kernel where the regularization parameter is set to 1 
and the epsilon is set to 0.2. 
The MLP model consists of an input layer, an output layer, 
and two simple hidden layers each of which has 1024 neurons.
The PCNN model, which was originally designed for inferring 
vector magnetic fields, 
is modified to output the LOS velocity and Doppler width,
though the same model architecture and hyperparameter setting are used here.

\begin{figure}[ht]
\epsscale{1.1}
\plotone{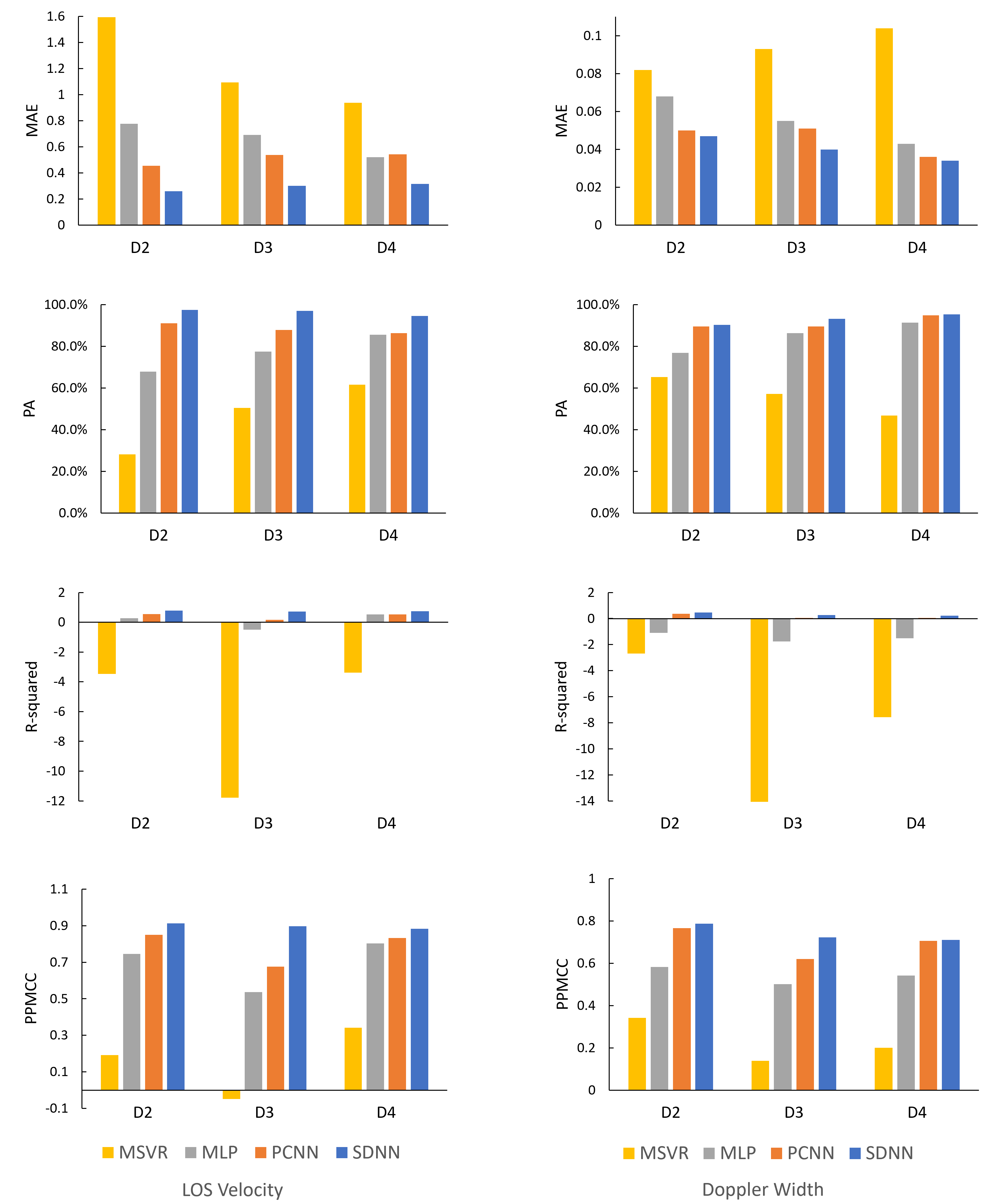}
\caption{Comparison of MSVR, MLP, PCNN and SDNN
based on the test image/dataset D2 
(D3, D4, respectively) from AR 12371
(AR 12665, AR 12673, respectively) collected on
2015 June 25 16:55:13 UT
(2017 July 13 16:20:12 UT,
2017 September 6 19:17:53 UT, respectively)
where training data were taken from D1 $\cup$ D5.
Left column: performance metric values, displayed by bar charts, that are obtained by the four machine learning (ML) methods in predicting LOS velocities.
Right column: performance metric values obtained
by the four ML methods in predicting Doppler widths.}
\label{fig:table_chart}
\end{figure}

Figure \ref{fig:table_chart} compares the MAE, PA, R-squared and PPMCC values
of the four ML methods based on the test image/dataset D2 
(D3, D4, respectively) from AR 12371
(AR 12665, AR 12673, respectively) collected on
2015 June 25 16:55:13 UT
(2017 July 13 16:20:12 UT,
2017 September 6 19:17:53 UT, respectively)
where training data were taken from D1 $\cup$ D5.
It can be seen from Figure \ref{fig:table_chart} that
SDNN outperforms the other three ML methods 
in terms of all the four performance metrics. 
When compared to the most closely related PCNN method,
SDNN achieves an MAE of
0.260 (0.302, 0.316, respectively) while
PCNN achieves an MAE of
0.455 (0.537, 0.542, respectively),
showing an improvement of
42.9\% (43.8\%, 41.7\%, respectively) by SDNN,
on D2 (D3, D4, respectively) 
in producing the LOS velocity maps for the three datasets.
Furthermore, in producing these LOS velocity maps,
SDNN achieves a PPMCC of
0.912 (0.896, 0.883, respectively) while 
PCNN achieves a PPMCC of
0.849 (0.676, 0.832, respectively) 
on D2 (D3, D4, respectively);
SDNN beats PCNN by
7.4\%, 32.5\% and 6\% respectively on the three datasets.
These results indicate that our SDNN model is more robust and has a better
generalization and inference capability
than the closely related PCNN model. 
It is worth pointing out that 
the granular patterns
in the LOS velocity map of the SDNN method
shown in Figure \ref{fig:granular_penumbral},
which exhibit important kinematic information on the photospheric surface \citep{Jaeggli_2016},
are missing from the LOS velocity maps produced by the other three ML methods,
another sign indicating the superiority of the proposed SDNN model.

\subsection{Comparison between the Inversion Results of SDO/HMI and GST/NIRIS}

So far we have focused on Stokes inversion for GST/NIRIS.
There are inversion results from
the Helioseismic and Magnetic Imager \citep[HMI;][]{2012SoPh..275..207S}
on board the Solar Dynamics Observatory \cite[SDO;][]{2012SoPh..275....3P}.
Here we compare the inversion results of the
space-borne observatory (SDO/HMI) and ground-based observatory (BBSO/GST).
We selected an HMI Dopplergram from AR 12673 collected 
on 2017 September 6 19:00:00 UT, and a temporally closest test image/dataset, denoted D6,
from the same AR 12673 collected by GST/NIRIS on 2017 September 6 19:01:48 UT.
We aligned the HMI and GST/NIRIS images, and
applied our SDNN model trained by the dataset 
D1 $\cup$ D5 and the ME inversion code 
described in Section \ref{sec:observational data}
to D6 respectively.

\begin{figure}[ht]
\epsscale{1.15}
\plotone{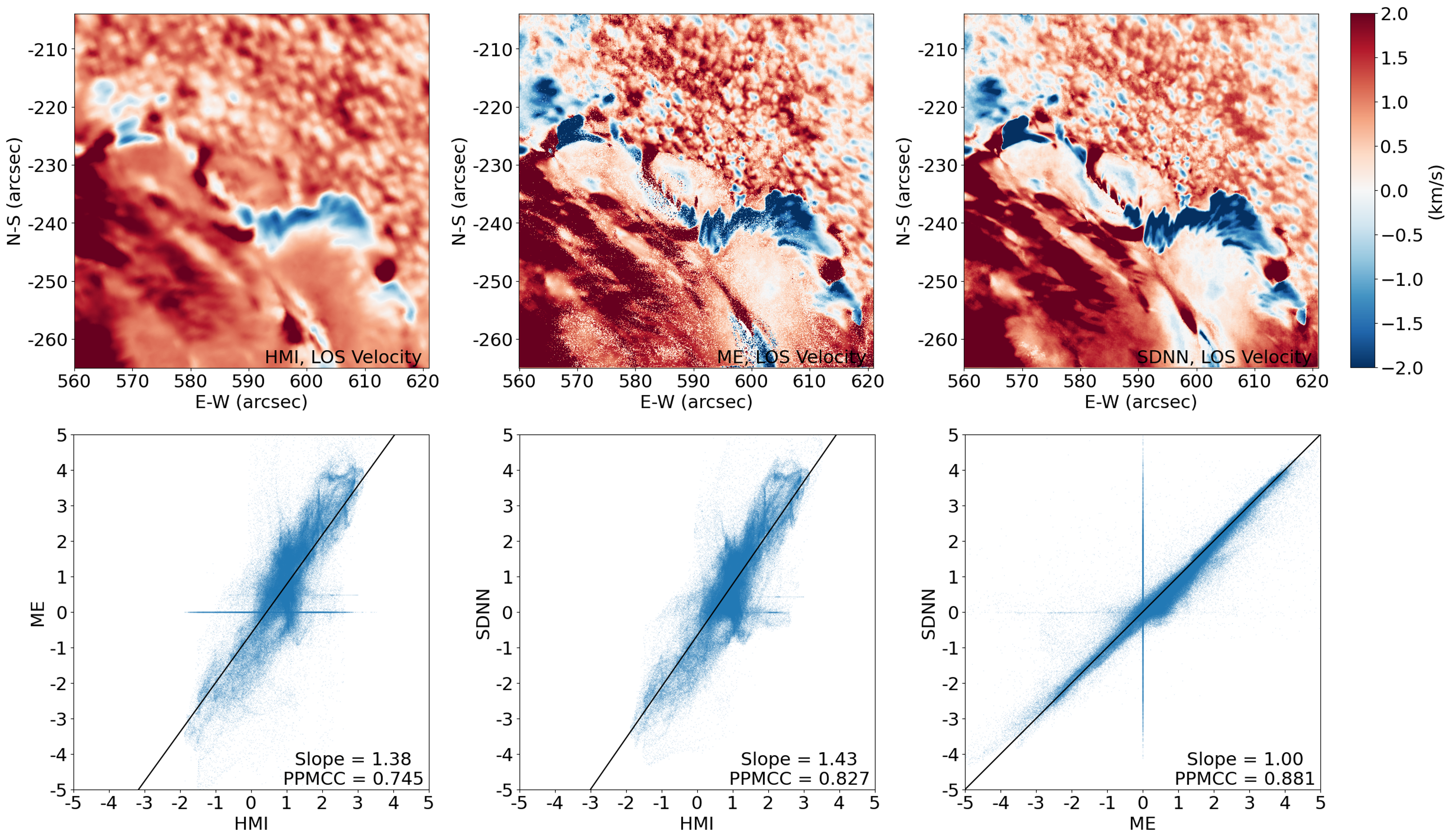}
\caption{Comparison between the inversion results of SDO/HMI and GST/NIRIS in AR 12673.
Top panels (from left to right): the LOS velocity map from 
the HMI Dopplergram collected on 2017 September 6 19:00:00 UT,
the LOS velocity map produced by the ME inversion code, and the LOS velocity map produced by SDNN 
with training data from D1 $\cup$ D5
where the ME code and SDNN were applied to the test image/dataset D6 collected by GST/NIRIS on 2017 September 6 19:01:48 UT.
Bottom panels (from left to right): the scatter plot between ME and HMI,
the scatter plot between SDNN and HMI, and the scatter plot between SDNN and ME.}
\label{fig:NIRIS_vs_HMI}
\end{figure}

Figure \ref{fig:NIRIS_vs_HMI} presents the LOS velocity map from the 
HMI Dopplergram and the LOS velocity maps produced by ME and SDNN
(top panels), and shows the corresponding scatter plots (bottom panels).
It can be seen from the top panels that 
the maps of GST/NIRIS obtained by ME and SDNN are much clearer than
the map from HMI. 
This happens due to the higher resolution imaging data offered by GST/NIRIS.
On the other hand, the maps from HMI and SDNN are smoother than the map of ME which contains
salt-and-pepper noise pixels.
Furthermore, we see from the left and middle scatter plots
in Figure \ref{fig:NIRIS_vs_HMI}
that the map produced by SDNN is closer to the map from HMI with a PPMCC of 0.827
than the map produced by ME which has a PPMCC of 0.745.
The black regression lines in the scatter plots have a slope of $\sim$1.4, possibly caused by the offset in calibration,
which occurs due to the intrinsic difference between 
the two instruments HMI and GST/NIRIS. 
These results further demonstrate that machine learning can complement the ME approach.
Specifically, our SDNN model is trained with the labels 
(i.e., LOS velocities and Doppler widths)
produced by the ME inversion code described in
Section \ref{sec:observational data}.
With the generalization and inference capabilities, 
the machine learning-based SDNN model 
can produce better inversion results than the calculation-based ME code.

\subsection{Extending the SDNN Method to Infer Vector Magnetic Fields}
\label{sec:extended results}

\begin{figure}[ht]
\epsscale{1.15}
\plotone{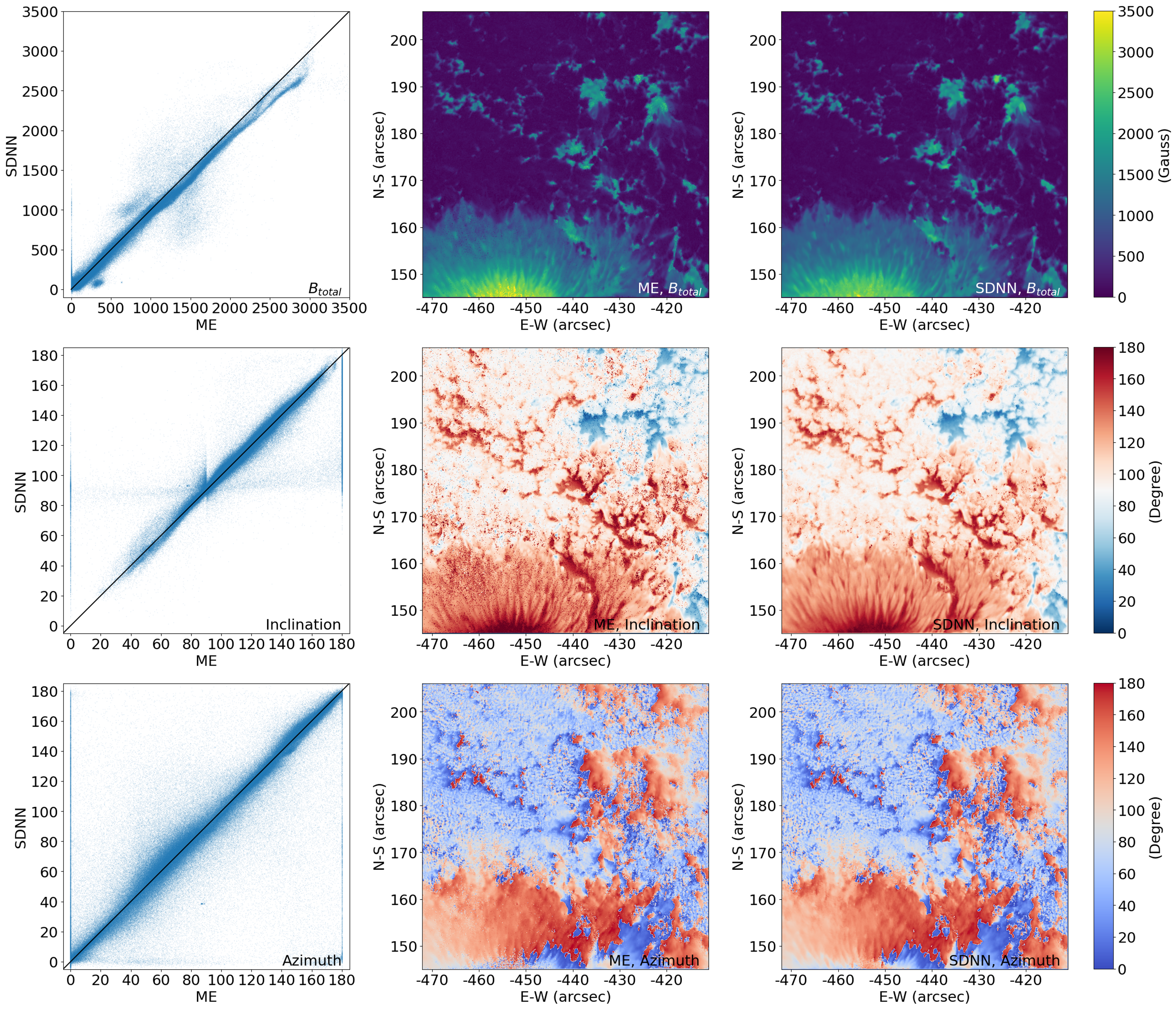}
\caption{Comparison between the ME inversion code and SDNN for producing the
total magnetic field strengths (top panels), 
inclination angles (middle panels), 
and azimuth angles (bottom panels) 
of pixels
based on the test image D3 from AR 12665 collected on 2017 July 13 16:20:12 UT, where training data were taken from D1 $\cup$ D5. The first column shows scatter plots where the X-axis and Y-axis represent the values obtained by the ME code and SDNN respectively.
	Pixels on the diagonal line in a scatter plot 
    have identical ME-calculated and SDNN-inferred values.
	The second and third columns show the magnetic field maps produced by the ME code and SDNN respectively.}
\label{fig:magnetic_field}
\end{figure}

Finally we extend our SDNN model to infer vector magnetic fields from
Stokes profiles of GST/NIRIS.
We normalized the total magnetic field strength ($B_{total}$), 
inclination angle and azimuth angle as done in \citet{Liu_2020}. 
The SDNN model was trained by the data in D1 $\cup$ D5. 
The training labels (i.e., 
total magnetic field strengths,
inclination angles and
azimuth angles) were produced by the
ME inversion code described in Section \ref{sec:observational data}.
Figure \ref{fig:magnetic_field}
presents inversion results 
obtained from the ME code and SDNN respectively,
based on the test set (image) D3 with data samples (pixels) from
AR 12665
collected on 2017 July 13 16:20:12 UT. 
(The inversion results for the test sets D2 and D4 are similar and omitted here.)
The first column in Figure \ref{fig:magnetic_field} shows scatter plots, 
the second column shows the magnetic field maps produced by the ME code
and the third column shows the magnetic field maps produced by SDNN.
Pixels on the diagonal line in a scatter plot 
have identical ME-calculated and SDNN-inferred values.

It can be seen from Figure \ref{fig:magnetic_field}
that the magnetic field maps 
produced by the ME code and SDNN are highly correlated,
with PPMCC being 0.980 
(0.867, 0.866, respectively)
for the total magnetic field strength (inclination angle, azimuth angle, respectively).   
Like LOS velocity maps and Doppler width maps,
the magnetic field maps produced by SDNN are smoother and cleaner than those
produced by the ME code.
We see many salt-and-pepper noise pixels in the magnetic field maps produced by the ME code.
When compared to the closely related PCNN method developed by \citet{Liu_2020},
SDNN achieves an MAE of
63.715 (5.721, 11.402, respectively) 
while PCNN achieves an MAE of
79.089 (6.270, 12.785, respectively), 
showing an improvement of
19.4\% (8.8\%, 10.8\%, respectively) by SDNN,
for the total magnetic field strength (inclination angle, azimuth angle, respectively).
Furthermore, SDNN outperforms PCNN in PPMCC
with SDNN's PPMCC being 
0.980 (0.867, 0.866, respectively) 
compared to 
PCNN's PPMCC of
0.976 (0.860, 0.848, respectively) 
for the total magnetic field strength (inclination angle, azimuth angle, respectively).

\section{Discussion and Conclusions} 
\label{sec:conclusion}

We develop a deep learning model (SDNN) to infer LOS velocities and Doppler widths
from Stokes profiles of GST/NIRIS at BBSO.
The labels for training SDNN are prepared by a Milne–Eddington (ME) inversion code
used by BBSO.
We compare the LOS velocity and Doppler width maps produced by SDNN with
those from the ME inversion code and related machine learning (ML) algorithms 
including 
multiple support vector regression (MSVR), 
multilayer perceptrons (MLP) 
and a pixel-level convolutional neural network (PCNN).  
We next compare the inversion results of the ME code and SDNN based on GST/NIRIS with
those from SDO/HMI in flare-prolific AR 12673.
Finally, we extend SDNN to infer vector magnetic fields
from Stokes profiles of GST/NIRIS.

Our main results are summarized as follows.
\begin{quote}
        1. For the test sets from GST/NIRIS considered in the paper, SDNN produces smoother and cleaner LOS velocity and Doppler width maps than the ME inversion code. 
        The same conclusion is obtained when comparing the inversion results of GST/NIRIS and SDO/HMI.
        Furthermore, SDNN performs Stokes inversion through making predictions, and hence is faster than the
        computation-based ME code. It takes $\sim$75 s for SDNN to process a test image here, which is approximately five times faster than the ME code.
        
        2. The SDNN-inferred LOS velocities are highly correlated to the ME-calculated ones with PPMCC being close to 0.9 on average. 
        The LOS velocity and Doppler width maps
        produced by SDNN are closer to ME’s maps than those from the related ML algorithms
        (MSVR, MLP, PCNN). Furthermore, both ME and SDNN are able to infer granular patterns in LOS velocity maps.
        These patterns exhibit important kinematic information on the photospheric surface,
        which are missing from the LOS velocity maps produced by the related ML algorithms.
        These results demonstrate the better learning capability of SDNN than the related ML algorithms.
    
        3. Training data has a significant impact on the performance of SDNN.
        When SDNN is trained by data from a normal active region (AR 12371),
        it performs well on normal active regions such as 
        AR 12665, but suffers on AR 12673, which is the most flare-productive active region in solar cycle 24 \citep{Wang_2018_inversion}.
        To acquire knowledge concerning the special pixels 
        in the extremely flare-productive active region AR 12673,
     we have to include some of the special pixels/data samples in the training set. 
        This finding is consistent with the guidelines suggested in the
        machine learning (ML) literature \citep{10.5555/1162264} where an iterative training process with increasing training data is often used to improve the performance of an ML model. 
        
        4. Our work focuses on LOS velocity and Doppler width inference. 
        The input of our SDNN method consists of
        Stokes \textit{I}, \textit{Q}, \textit{U}, \textit{V} profiles of GST/NIRIS.
        One wonders whether using Stokes \textit{I} suffices to infer the
        LOS velocity and Doppler width.
        We have conducted additional experiments in which SDNN is merely trained by
        Stokes \textit{I}
        of pixels in D1 $\cup$ D5.
        Our experimental results showed that 
        using the four Stokes \textit{I}, \textit{Q}, \textit{U}, \textit{V} profiles
        performs better than using Stokes \textit{I} only.
        Specifically, with only Stokes \textit{I} as the input, SDNN
        obtains an MAE of 0.774 (0.395, 0.534, respectively) and PPMCC of 0.412 (0.782, 0.787, respectively) for
        LOS velocity inference,
        compared to the MAE of 0.260 (0.302, 0.316, respectively) and PPMCC of 0.912 (0.896, 0.883, respectively)
        obtained by the four Stokes profiles, 
        showing a deterioration of 197.7\% (30.8\%, 69.0\%, respectively) on MAE and 54.8\% (12.7\%, 10.9\%, respectively) on PPMCC 
        in D2 (D3, D4, respectively). 
        Similarly, with only Stokes \textit{I} as the input, SDNN obtains
        an MAE of 0.093 (0.059, 0.047, respectively) and PPMCC of 0.010 (0.343, 0.441, respectively) for
        Doppler width inference, 
        compared to the MAE of 0.047 (0.040, 0.034, respectively) and PPMCC of 0.787 (0.723, 0.711, respectively)
        obtained by the four Stokes profiles, 
        showing a deterioration of 97.9\% (47.5\%,  38.2\%, respectively) on MAE and 98.7\% (52.6\%, 38.0\%, respectively) on PPMCC
        in D2 (D3, D4, respectively).
        These results indicate that one should use all the four Stokes profiles of GST/NIRIS
        for LOS velocity and Doppler width inference.
        
        5. SDNN performs equally well for vector magnetic field inference. 
        For example, with the four Stokes \textit{I}, \textit{Q}, \textit{U}, \textit{V}
        as the input, D1 $\cup$ D5 as the training set and D3 as the test set,
        SDNN achieves an MAE of
        63.715 (5.721, 11.402, respectively) 
        while the most closely related PCNN method developed by \citet{Liu_2020} achieves an MAE of
        79.089 (6.270, 12.785, respectively), 
        showing an improvement of
        19.4\% (8.8\%, 10.8\%, respectively) by SDNN,
        for the total magnetic field strength (inclination angle, azimuth angle, respectively).
        Furthermore, SDNN's PPMCC values are
        0.980 (0.867, 0.866, respectively),
        which are better than
        PCNN's PPMCC values of
        0.976 (0.860, 0.848, respectively) 
        for the total magnetic field strength (inclination angle, azimuth angle, respectively).
        Similar results are obtained for the test sets D2 and D4.
\end{quote}

Based on these results, we conclude that SDNN 
is an effective and alternative method for 
inferring LOS velocities, Doppler widths
and vector magnetic fields
from Stokes profiles of GST/NIRIS.
It is hoped that 
SDNN will be a useful tool in producing 
high-quality velocity and magnetic fields that are crucial for understanding 
the evolution of physical properties of the solar atmosphere.
\  \\

The authors gratefully acknowledge the use of data from the Goode Solar Telescope (GST) of the Big Bear Solar Observatory (BBSO). 
The BBSO operation is supported by NJIT and U.S. NSF grant AGS-1821294. 
The GST operation is partly supported by the Korea Astronomy and Space Science Institute and the Seoul National University. 
This work was supported by U.S. NSF grants AGS-1927578, AGS-1954737 and AGS-2228996. 
K.A. and W.C. acknowledge the support of NASA under grant 80NSSC20K0025. 
Q.L., Y.X. and H.W. acknowledge the support of NASA under grants
80NSSC18K1705, 80NSSC19K0068 and 80NSSC20K1282.
We thank the scientific editor
for his guidance and an
anonymous referee for the thoughtful comments that have helped us
improve the presentation and content of this paper.

\vspace{5mm}
\facilities{Big Bear Solar Observatory, Solar Dynamics Observatory}

\bibliographystyle{aasjournal}

\end{document}